\documentclass[journal]{IEEEtran}
\ifCLASSINFOpdf
\usepackage[pdftex]{graphicx}
 \usepackage{epstopdf}
\usepackage{color}
\usepackage{transparent}

\else
\fi
\usepackage[cmex10]{amsmath}

\usepackage{cases}
\usepackage{amssymb}
\ifCLASSOPTIONcompsoc
  \usepackage[caption=false,font=normalsize,labelfont=sf,textfont=sf]{subfig}
\else
  \usepackage[caption=false,font=footnotesize]{subfig}
\fi
\begin{document}
%
\title{Nyquist Filter Design using POCS Methods: Including Constraints in Design}
%
%
%

\author{Sanjeel~Parekh~\IEEEmembership{}and~Pratik~Shah~\IEEEmembership{}
\thanks{Sanjeel Parekh is a B.Tech student at The LNM Institute of Information Technology, Jaipur,
India, 302031  e-mail:sanjeelparekh@gmail.com.} 
\thanks{Pratik Shah received his Ph.D from DAIICT, Gandhinagar, India and is currently working with The LNM Institute of Information Technology, Jaipur as an Assistant Professor e-mail:pratik.shah@lnmiit.ac.in.} }

%
%

\markboth{}%
{Shell \MakeLowercase{\textit{et al.}}: Bare Demo of IEEEtran.cls for Journals}
%



\maketitle




\begin{abstract}
The problem of constrained finite impulse response (FIR) filter design is central to signal processing and arises in a variety of disciplines. This paper surveys the design of such filters using Projection onto convex sets (POCS) and discusses certain commonly encountered time and frequency domain constraints. We study in particular the design of Nyquist filters and propose a simple extension to the work carried out by Haddad, Stark, and Galatsanos in \cite{861404}. The flexibility and the ease that this design method provides in terms of accommodating constraints is one of its outstanding features.
\end{abstract}


%
\IEEEpeerreviewmaketitle

\section{Introduction}
%
%
%
%


\IEEEPARstart{F}{IR} filter design is one of the most basic and important problems in digital signal processing. A classical filter design problem imposes constraints on the passband, transition band and stopband fluctuations and the number of filter coefficients. In order to widen the applicability of FIR filters it was necessary to devise methods to efficiently impose additional or ingenious constraints. For example, filters with linear phase constraints are very efficiently implemented using method proposed by McClellan-Parks in \cite {1083764} but imposition of additional constraints is not possible.\\  \\ If the constraints on a filter can be formulated as  closed convex sets, where the intersection of the sets is assumed to be non-empty, the POCS methods serve as a very powerful design tool. After relaxation of certain conditions good results are obtained for non-convex sets too see \cite {861404}. In \cite{5350241} POCS methods have been used for reducing the number of non-zero coefficients. These methods have been easily and  successfully extended to the design of multi-dimensional filters, as demostrated in \cite{581378,1085572}. This further emphasizes the usefulness of the POCS methods. It is worth mentioning here that \textit{POCS methods guarantee a feasible solution, not an optimal one.} \\ \\ This paper surveys the technique of constrained FIR filter design using projections onto convex sets with particular emphasis on constraints presented in \cite {861404}. In \cite{581378}, a FFT based iterative algorithm was proposed, which turned out to be a specific case of the work done in \cite {861404}. \\ \\ We demonstrate how the constraints are formulated for given design problem.  We also provide an implementation of constrained FIR filter design using POCS methods. As an extension to the design method stated in \cite{861404} we consider design of a very important class of filters known as the Nyquist filters. 

\noindent The design of  Nyquist filters introduces certain zero valued coefficients in the impulse response, which makes them computationally more efficient than low pass filters of the same order. These filters find applications in perfect reconstruction filter banks, interpolator and decimator design and non-uniform sampling, see \cite{mitra2006digital,1452570}. In particular, the half band filters are used to design the Hilbert transformer also, see \cite{mitra2006digital, hbfdsp}.

\section{Preliminaries}

We begin with a review of the fundamental theorem of POCS and the classical linear phase FIR filter problem. 

In a Hilbert space $H$, if we consider a closed convex set $C$, then the projection of an arbitrary vector $x \in H$ onto $C$ is denoted by $x^*=Px$ and its defined by the minimum distance point in $C$ from $x$. The projection of x on the set C is defined as:

\begin{equation}
\label{eqn1}
\|x-x^*\| =\min _{g \in C}\|x-g\|
\end{equation}
where $P$ is the projector for set C. The fundamental theorem of POCS states that given $n$ closed convex sets $ C_1, C_2, \ldots, C_n $ in a Hilbert space $H$, let $P_1, P_2, \ldots, P_n$ be the projectors associated with these sets respectively, then the iterative procedure given by

\begin{equation}
\label{eqn_example}
x^{k+1} = P_{m} P_{m-1}, \ldots P_{1}x^k
\end{equation}

\noindent converges weakly to a point in $C_0$ where $C_0$ is defined as $C_0=\bigcap _{i=1}^{n} C_i$. This theorem, is in general true even for relaxed projectors defined by $T_i \equiv I+\mu_i(P_i-I)$ where $I$ is the identity operator and $0< \mu_i <2$ and $i=1, 2, \ldots, n$.

For typical filter design problems, $\mathbb{R}^M$ is an appropriate Hilbert space. Let $\mathbf{x},\mathbf{y} \in \mathbb{R}^M$, then the inner product is defined as:
\[\langle x,y \rangle=\sum_{n=0}^{M-1}x(n)y(n).\]

This inner product induces standard 2-norm on the euclidean space  $\mathbb{R}^M$. \\

Before proceeding further let us define the basic linear phase FIR filter design problem that we would be considering in subsequent sections with additional constraints.\\

\noindent An FIR filter is characterized by its system function given by

\begin{equation}
H(z)=\sum_{n=0}^{N-1}h(n)z^{-n}
\end{equation}

\noindent where $N$ is the filter length and $ h(0), h(1), \ldots, h(N-1) $ filter coefficients. For  linear phase, the filter coefficients must satisfy the following condition:

\begin{equation}
h(n)=\pm h(N-1-n),\text{ for }n=0,1, \ldots,N-1
\end{equation}

\noindent An ideal low pass filter frequency response would require unity magnitude in passband, zero in stopband and a sharp cutoff from passband to stopband. Since it is impossible to meet criteria of an ideal filter, in general a filter design problem has the following specifications:
\begin{enumerate}
\item{Maximum tolerable passband ripple, denoted by $\alpha$}.
\item{Maximum tolerable stopband ripple, denoted by $\beta$}.
\item{Passband edge frequency, denoted by $\omega_p$}.
\item{Stopband edge frequency, denoted by $\omega_s$}.
\end{enumerate}

 If $A(\omega)\equiv|H(\omega)|$ and $\phi(\omega)\equiv \angle H(\omega)$, then the linear phase FIR filter design problem can be mathematically formulated as

\begin{itemize}
\item{$A(\omega) \in [1-\alpha,1+\alpha ]$ and $\phi(\omega)=\frac{-\omega(N-1)}{2}$\\ for $ \omega \in \Omega_p \equiv (\omega:0\leq \omega\leq\omega_p)$ }
\item{$A(\omega)\leq \beta$ for $ \omega \in \Omega_s \equiv (\omega:\omega_s\leq \omega\leq\pi)$ }
\end{itemize}


\subsection{Organization of the Article}
First we will discuss the various exisitng convex set formulations for the filter design problem. It is easy to extend each of them to higher dimensions. Next we discuss several important time and frequency domain constraints, with particular attention to Nyquist filters. In order to deal with linear phase FIR Nyquist filter design, we propose a simple extension of design methodology given in \cite{861404}. Finally, we show results of certain constrained FIR filters designed in MATLAB.

 Each of the sections would reiterate the importance of the POCS methods in imposing a new constraint on FIR filters, assuming that the constraint gives rise to a closed and convex set.


\section{Convex set formulation}
In this section we consider two convex set formulations. The Abo-Taleb and Fahmy formulation in \cite{1085572} was given in 1984 for 2-D filter design and the one discussed in  \cite{861404} was put forward in 2000.

\subsection{}
For completeness, we begin with the formulation presented in \cite{861404}. The Hilbert space under consideration is $\mathbb{R}^M$ where $M \gg N$ for high resolution Fourier transform. We define the following constraint sets:

\begin{eqnarray}
C_1 &\equiv& \{\mathbf{h} \in \mathbb{R}^M: h(n)=h(N-1-n),  \nonumber \\ && \text{for } n=0,1, \ldots, N-1 \} \nonumber \\
C_2&\equiv& \{\mathbf{h} \in \mathbb{R}^M: 1-\alpha\leq A( \omega)\leq 1+\alpha,  \nonumber \\ && \phi(\omega)=\frac{-\omega(N-1)}{2},   \text{ for }  \omega \in \Omega_p\} \nonumber \\
C_3&\equiv& \{\mathbf{h} \in \mathbb{R}^M: A( \omega)\leq \beta, \text{ for }\omega \in \Omega_s\}\nonumber \\
\end{eqnarray}

The set $C_1$ captures the linear phase property of the filter. Here the filter length $N$ is assumed to be odd. $C_2$ and $C_3$ describe the passband and the stopband requirements respectively with the desired tolerance values $\alpha$ and $\beta$.

We now check for their convexity, closedness and derive projectors for each one of them.\\

\subsubsection{Convexity of $C_1$}
Let $\mathbf{h_1,h_2} \in C_1$ then for $t\in[0,1]$
\begin{eqnarray}
h_3(n) &=& th_1(n)+(1-t)h_2(n) \nonumber \\
&=&  th_1(N-n-1)+(1-t)h_2(N-n-1) \nonumber\\
&=& h_3(N-n-1) \nonumber
\end{eqnarray}

Hence $C_1$ is convex.\\
\subsubsection*{Closedness of $C_1$}

Let the sequence $\{\mathbf{h_k}\} \rightarrow \mathbf{h^*} \text{ as } k \rightarrow \infty$. By definition, $\sum_{n=0}^{N-1} |h_k(n)-h^*(n)|^2 \rightarrow 0 \Rightarrow h_k(n) \rightarrow h^*(n)$. Since $ \mathbf{h_k} \in C_1,  h_k(N-1-n) \rightarrow h^*(n) $. Thus, $ h^*(n)=h^*(N-n-1)$.\\

\subsubsection*{Projector of $C_1$}
Let $\mathbf{g} $ be an arbitrary vector in $\mathbb{R}^M$, then its projection $\mathbf{h^*}$  is given by:

\begin{numcases}{ h^*(n)=}
\label{star11}
\frac{g(n)+g(N-1-n)}{2} \text{, for }  n = 0, 1, \ldots, N-1  \nonumber\\
    0 \text{, elsewhere}
\end{numcases}



The above result is very intutive. Since the set $C_1$ is symmetric, in order to minimize the norm, any arbitrary vector that is to be projected must also be symmetric, which is nothing but the mathematical requirement expressed in (\ref{star11}).\\
\subsubsection{Convexity and Closedness of $C_2$}
Let $\mathbf{h_1, h_2} \in C_2$, and $ \mathbf{h_3}=t \mathbf{h_1} +(1-t)\mathbf{h_2} \Rightarrow  H_3(\omega) =t H_1(\omega) +(1-t)H_2(\omega)$ for $\omega \in \Omega_p$. Since $H_1(\omega)=A_1(\omega) \exp ^{j \phi(\omega)}$ and $H_2(\omega)=A_2(\omega) \exp ^{j \phi(\omega)}$, the phase of $H_3(\omega)$ is also $ \phi(\omega)$ and the lower and upper bound on magnitude implies that $A_3(\omega) =t A_1(\omega) +(1-t)A_2(\omega) \in [1-\alpha, 1+\alpha]$. Thus, $C_2$ is convex.

It must be noted that to check for convexity of $C_2 $, the phase along with the magnitude is to be considered. Without the phase constraint it is possible to choose two vectors $\mathbf{h_1,h_2} \in C_2$ with values of $\phi$ which violate the lower bound on the  magnitude, thus violating convexity. Since $C_2$ includes all its limit points it is closed.\\
\subsubsection*{Projector of $C_2$}

\begin{figure}[htb]
\def\svgwidth{\columnwidth}
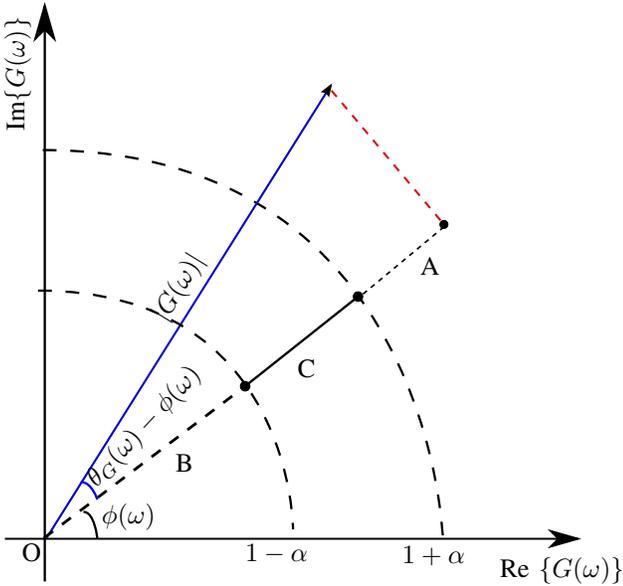
 \caption{Geometric representation of $C_2$ and its projector}
\label{c2}
\end{figure}

We deduce the projector from Fig.\ref{c2}. The figure is drawn for a particular $\omega \in \Omega_p$. First the point $G(\omega)$ is projected onto the line with slope $\phi (\omega)$ and then appropriate decision is taken for the magnitude of the projector. For $\omega \not\in \Omega_p$ the point is left unchanged. Thus if we denote the projector by $h^*$ and its Fourier transform by $h^*(n) \leftrightarrow H^*(\omega)$  we have:

\begin{numcases}{H^*(\omega)=}
(1+ \alpha) e^{j \phi (\omega)} \text{,  if case A } \nonumber\\
(1- \alpha) e^{j \phi (\omega)} \text{,  if case B } \nonumber\\
|G(\omega)| \cos[\theta_G(\omega)-\phi (\omega)]e^{j \phi (\omega)} \text{,  if case C } \nonumber\\
G(\omega) \text{,  if } \omega \not\in \Omega_p 
\end{numcases}

where \\
Case A is: $|G(\omega)| \cos[\theta_G(\omega)-\phi (\omega)] \geq 1+\alpha$\\
Case B is: $|G(\omega)| \cos[\theta_G(\omega)-\phi (\omega)] \leq 1-\alpha$\\
Case C is: $1-\alpha \leq |G(\omega)| \cos[\theta_G(\omega)-\phi (\omega)] \leq 1+\alpha$\\ 

\subsubsection{Convexity and Closedness of $C_3$}
Let $\mathbf{h_1, h_2} \in C_3$, and $ \mathbf{h_3}=t \mathbf{h_1} +(1-t)\mathbf{h_2} \Rightarrow  H_3(\omega) =t H_1(\omega) +(1-t)H_2(\omega)$ for $\omega \in \Omega_s$. Using the fact that for any two complex numbers $A$ and $B$, $ |A+B| \leq |A| + |B|$ we have:

\begin{eqnarray}
| H_3(\omega)| &=&|t H_1(\omega) +(1-t)H_2(\omega)|\nonumber \\
& \leq &  t| H_1(\omega)| + (1-t) |H_2(\omega)|\nonumber\\
&\leq & \beta\nonumber
\end{eqnarray}
Thus $C_3$ is a convex set. Since $C_3$ includes all its limit points it is also closed.


\subsubsection*{Projector of $C_3$}

\begin{numcases}{H^*(\omega)= }
 \frac{\beta G(\omega)}{|G(\omega)|}  \text{, for }|G(\omega)|\geq \beta \nonumber\\
    G(\omega) \text{, elsewhere}
\end{numcases}

Just like in the case of $C_2$ the deduction of the projector is easily done from its geometric representation. This set, in the frequency domain for all $\omega \in \Omega_s$, would represent a circle with radius $\beta$. Clearly the distance of all the points with magnitude greater than $\beta$ would be minimized from the disk  if they are projected onto the boundary. The other points are left unchanged.

Thus, putting together all the sets the linear phase FIR filter design algorithm is given by: $\mathbf{h}^{k+1}= P_1P_2P_3 \mathbf{h}^k$ with arbitrary $\mathbf{h}^0$.


\subsection{}

Next, we consider the Abo-Taleb Fahmy algorithm suggested in \cite{1085572} for the one dimensional case. The discussion below assumes a symmetric odd length linear phase filter. The generalization to other types is straight forward as suggested in \cite{1083764}. The amplitude response of a linear phase FIR filter can be given by

\begin{equation}
\label{atf1}
H(\omega_k)=\sum_{n=0}^{(N-1)/2} a(n) \psi_{n}(\omega_k)
\end{equation}

where $N$ is the filter length, $\omega_k$ is the discrete set of $K$ frequencies in the passband and the stopband, $\psi_{n}(\omega_k) = \cos (2\pi n \omega_k)$ for $ n=0, 1, \ldots, (N-1)/2$ and $a(n)$ are related with the filter coefficients for an odd filter length, $N$ by the following relation:
\begin{equation}
a(0)=h(0), \text{ }  a(n)=2h\Big(\frac{N-1}{2}-n\Big)
\end{equation}

For a given desired frequency response $D(\omega_k)$ our aim is to constrain the error function $e(\omega_k)$ within certain limit $\lambda(\omega_k)$  i.e.

\begin{equation}
\label{atf2}
e(\omega_k)=|D(\omega_k)-H(\omega_k)| \leq \lambda(\omega_k)
\end{equation}

For a given filter length, $N$ and filter coefficients $h(0), h(1), \ldots , h(N-1)$, we define the above equation as a set of $K$ convex closed sets

\begin{equation}
\label{atf3}
C'_{k} \equiv \{\mathbf{a}:|D(\omega_k)-\sum_{n=0}^{(N-1)/2} a(n) \psi_{n}(\omega_k)| \leq \lambda(\omega_k)\}
\end{equation}

The iteration rule used in \cite{1085572} for finding a point in the intersection of all the $K$ sets is an interesting one. For each iteration, projections are made for that set which has the maximum error among all the frequencies. We give below the proofs for convexity and closedness for $C'_{k}$ and derive the projector for same.


\subsubsection{Convexity}

Let $\mathbf{a_1, a_2} \in C'_{k}$ and $H_1$ and $H_2$  be the corresponding responses satifying equation (\ref{atf1}). Then $a_3=\beta \mathbf{a_1} +(1-\beta)  \mathbf{a_2} $ and response $H_3$ due to $ \mathbf{a_3} $ satisfies $H_3= \beta H_1 +(1-\beta) H_2$. Therefore,

\begin{eqnarray}
|D-H_3|&{}={}&|D-\beta H_1 - (1-\beta) H_2|\nonumber\\
&{}={}&|\beta(D- H_1) + (1-\beta)(D- H_2)|\nonumber\\
&{}\leq{}&\beta|D- H_1| + (1-\beta)|D- H_2|\nonumber\\
&{}\leq{}&\beta \lambda + (1-\beta)\lambda =\lambda \nonumber\\
\end{eqnarray}

Since $ \mathbf{a_3}  \in C'_{k}$, it is convex.\\

\subsubsection{Closedness}

Let $\mathbf{a^*}$ be the limit point of a sequence $ \{\mathbf{a_n}\}$ in $C'_{k}$ as $ n \rightarrow \infty$. We prove that $\mathbf{a^*} \in C'_{k}$ using contradiction. Assume that $\mathbf{a^*} \not\in  C'_{k}$ i.e. $|D-H^*| =\lambda + \eta$. Since $ \eta > 0$, rearranging the equation we have

\begin{equation}
\label{atf4}
\eta=||D-H^*|-\lambda|
\end{equation}

Since $\mathbf{a_n} \in  C'_{k} \Rightarrow |D-H_n| \leq \lambda$. Substituting $|D-H_n|$ for $\lambda$ in (\ref{atf4}) we have

\begin{eqnarray}
\eta&{}\leq{}&| |D-H^*|-|D-H_n| |\nonumber\\
&{}\leq{}&| (D-H^*)- (D-H_n) |\nonumber\\
\Rightarrow {} \eta &{}\leq{}&| H_n - H^* |\nonumber\\
\end{eqnarray}

Since $H^*$ is the limit point, the right hand side of the above equation by definition goes to 0 as  $ n \rightarrow \infty$, implying that $\eta \leq 0$. Thus we arrive at a contradiction. Hence $\mathbf{a^*} \in  C'_{k}$, i.e. the set is closed.\\

\subsubsection{Projector}
The projection of each of the filter coefficients of an arbitrary vector $\mathbf{a}=[a(0), a(1), \ldots, a(n)]$ is given by
\begin{numcases}{a(n)^{i+1}=}
\label{atfp}
a(n)^i - [\lambda(\omega_k) - |e(\omega_k)|] \text{sign}(e_k) \nonumber\\ .\psi(\omega_k)/ \sum_{m=0}^{(N-1)/2} \psi_m^2 (\omega_k), \text{ if } e(\omega_k) > \lambda(\omega_k)
\nonumber\\
a(n)^i,  \text{ if } e(\omega_k) \leq \lambda(\omega_k)
\end{numcases}

Here we provide a sketch of the derivation. For details the reader is referred to \cite{1085572}. The Lagrange multiplier method is used where we minimize the expression given by:
\begin{equation}
M= \|\mathbf{a}^{i+1}-\mathbf{a}^{i}\|+ \alpha \Big(|D(\omega_k)-\sum_{n=0}^{(N-1)/2} a(n)^{i+1} \psi_{n}(\omega_k)| -\lambda(\omega_k) \Big)
\end{equation}

Taking its derivative with respect to $a(n)^{i+1}$, and determining the value of $\alpha$, the lagrange multiplier for which the norm is minimized, we arrive at the projector given in (\ref{atfp}).

The algorithm is an intutive one, where,  in each iteration the set which violates the constraint set the most is projected onto $C'$ where, $C'$ is given by $C'=\bigcap _{k=1}^{K} C'_k$.


\section{Additional Constraints}

A situation where we want for a particular input $\mathbf{s}$, the output of a filter to be restricted within certain limits arises in many signal processing applications. Such a constraint is both closed and convex, thus can be implemented using the POCS methods. Mathematically, it can be formulated as follows
\begin{eqnarray}
C_4  \equiv \{\mathbf{h} \in \mathbb{R}^M : b_1(n) \leq (\mathbf{s} * \mathbf{h})_n\leq  b_2(n),\nonumber \\  h(n)=0, n>N-1\}
\end{eqnarray}

Here $n=0,1,\ldots N+L-2, \text{ and } L$  denotes the length of the input signal. The convolution output can also be represented in the matrix form as $\hat{y}=S\mathbf{h}$, where $S$ is a $N+L-1 \times N$ matrix. If we denote the $n^{th}$ row of $S$ by $\mathbf{s_n}$, then it follows that $\hat{y}(n)=\mathbf{s_{n}}^T\mathbf{h}$. Here $\hat{y}, \mathbf{s_n}, \mathbf{h}$ are assumed to be column vectors. Thus $C_4$ can be written as:

\begin{eqnarray}
\label{cons4}
C_4 \equiv \{\mathbf{h} \in \mathbb{R}^M : b_1(n) \leq \mathbf{s_{n}}^{T}\mathbf{h} \leq  b_2(n),\nonumber \\ \text{ for } n=0,1,\ldots N+L-2\}
\end{eqnarray}

Here $\mathbf{s_{n}}^{T}\mathbf{h}$ is the inner product of $\mathbf{s_n}$ and $\mathbf{h}$ which is referred to as the \emph{soft linear constraint} in \cite{Stark:1998:VSP:551275}. We provide here the expression for projection. Projector of an arbitrary vector $\mathbf{g}$ for $n=0,1, \ldots N-1$ is given by

\begin{numcases}{h^*=P_4(\mathbf{g})=}
\label{proj4}
g \text{, if }  b_1(n) \leq \langle s_{n}^T,g\rangle \leq  b_2(n) \nonumber\\
x+ \frac{b_1(n)- \langle s_{n}^T,g\rangle}{\|s_n\|^2}s_n \text{, if}  \langle s_{n}^T,g\rangle <  b_1(n) \nonumber\\
x+ \frac{b_2(n)- \langle s_{n}^T,g\rangle}{\|s_n\|^2}s_n  \text{, if} \langle s_{n}^T,g\rangle >  b_2(n) \nonumber\\
\end{numcases}
The design algorithm is given by: $\mathbf{h}^{k+1}= P_1P_2P_3P_4\mathbf{h}^k$, with arbitrary $\mathbf{h}^0$.

For a particular input $\mathbf{s}$ we might want to constrain the output energy within certain limits. This would be a quadratic constraint, and can be modelled as
\begin{equation}
C_5 \equiv \{\mathbf{h} \in \mathbb{R^M} : \| (\mathbf{s} *\mathbf{ h})-d\| \leq \sigma  \}
\end{equation}
Here again we represent convolution in the martix form i.e. $S\mathbf{h}$. This  set would be an ellipsoid and thus both closed and convex. The projector is determined by finding the minimum of the lagrange function $J(h)$ given by:
\[J(h)=(g-h)^T(g-h)+\lambda((Sh-d)^T(Sh-d)-\sigma^2)\]
where $\lambda$ is the lagrange multiplier. Equating the derivative of $J$ with resect to  $h$ to zero we get:

\begin{eqnarray}
0&=&-2(g-h)^T+2\lambda (Sh-d)^TS  \nonumber\\
(g-h)^T&=&\lambda (Sh-d)^TS  \nonumber\\
(g-h)&=&\lambda S^T(Sh-d)  \nonumber\\
h_\lambda&=&(I+\lambda S^TS)^{-1}(g+\lambda S^Td)
\end{eqnarray}

So the projection of $\mathbf{g}$ for $ n=0,1, \ldots N-1$ is
\begin{numcases}{h^*=P_5(\mathbf{g})=}
\label{proj5}
 g \text{, if } \|(Sg-d)\| \leq \sigma \nonumber\\
 h_\lambda  \text{, if } \|(Sg-d)\| > \sigma
\end{numcases}
 The computation of $\lambda$ can be  done using Newton-Raphson method.
The design algorithm is given by: $\mathbf{h}^{k+1}= P_1P_2P_3P_5\mathbf{h}^k$, with arbitrary $h^0$.


\subsection{Arbitrary Phase and Magnitude Constraint}
Now we explore the problem of controlling both phase and magnitude at specific frequencies. So we must define our set formulation again to achieve the desired results. For a particular $\omega$ we require $|H(\omega)| \in [ a(\omega)-\delta,  a(\omega)+\delta]$ and $\phi(\omega) (\text{i.e.} \angle H(\omega)) \in [\alpha(\omega)-\epsilon, \alpha(\omega)+\epsilon] $ for arbitrary tolerances $\delta$ and $\epsilon$. Sets are defined as follow:

\begin{eqnarray*}
C_6 &\equiv&\{\mathbf{h} \in \mathbb{R}^M : a(\omega)-\delta \leq |H(\omega)| \nonumber \\ && \leq  a(\omega)+\delta \text{ and }  \alpha(\omega)-\epsilon \leq \phi(\omega) \leq  \alpha(\omega)+\epsilon \}\\
C_7 &\equiv& \{\mathbf{h} \in \mathbb{R}^M :h(N-1) \not= 0 \text{ and }\nonumber \\ && h(n)=0 \text{ for } n=N, N+1, \ldots,M-1\}
\end{eqnarray*}

The set $C_6$ is non convex (recall the importance of  linear phase constraint in making $C_2$ a convex set). So, as suggested in \cite{861404, Stark:1998:VSP:551275} summed distance error convergence has to be used to determine a point in the intersection of the sets under consideration. This may not be the case always but for this purpose it has shown good results \cite{861404} . On the other hand $C_7$ is a convex closed set with the projector for an arbitrary vector $\mathbf{g} \in \mathbb{R}^M$ given by:

\begin{numcases}{h^*=P_7(g)=}
\label{proj7}
 h^*(n)= g(n) \text{, for } 0 \leq n\leq N-1 \nonumber\\
 0 \text{, otherwise } 
\end{numcases}

\begin{figure}[htb]
\def\svgwidth{\columnwidth}
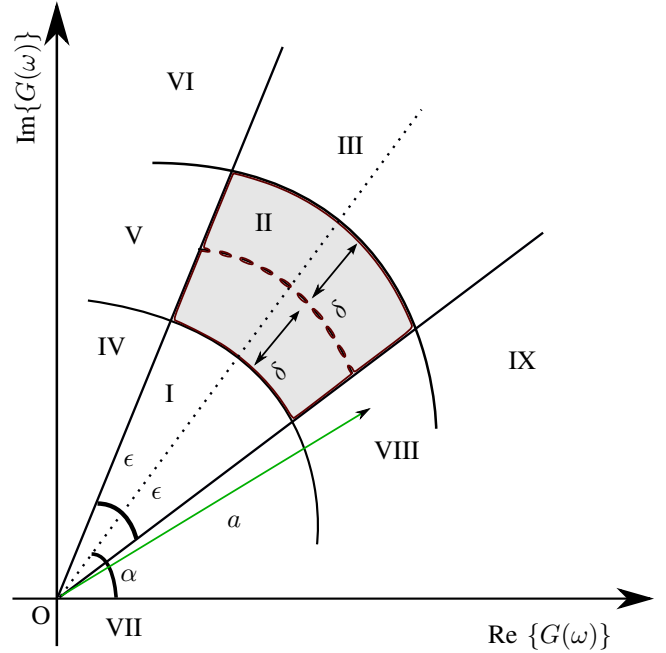
 \caption{Geometric representation of $C_6$ and its projector}
\label{c6}
\end{figure}

To compute the projector for set $C_6$ we analyse the geometry of the problem in frequency domain. Once we divide the problem into specific regions as done in Fig.\ref{c6}, the projection of an arbitrary vector $g \leftrightarrow G(\omega) \equiv x$  in each region is carried out according to Table \ref{table_c6}. Depending on the region the vector may satisfy both, one or none of the constraints of phase and magnitude. We project accordingly onto the shaded region of interest II.

\begin{table}[!t]
\renewcommand{\arraystretch}{1.3}
\caption{Projections for $C_6$}
\label{table_c6}
\centering
\begin{tabular}{c||c}
\hline
\bfseries Regions & \bfseries Projection\\
\hline\hline
I & $(a-\delta)\exp[j(\angle x)]$\\
II & $x$\\
III & $(a+\delta)\exp[j(\angle x)]$\\
IV & $(a-\delta)\exp[j(\alpha+\epsilon)]$\\
V & $|x|\cos(\angle x -\alpha-\epsilon)\exp[j(\alpha+\epsilon)]$\\
VI & $(a+\delta)\exp[j(\alpha+\epsilon)]$\\
VII & $(a-\delta)\exp[j(\alpha-\epsilon)]$\\
VIII & $|x|\cos(\alpha-\epsilon-\angle x)\exp[j(\alpha-\epsilon)]$\\
IX & $(a+\delta)\exp[j(\alpha-\epsilon)]$\\
\hline
\end{tabular}
\end{table}
The design algorithm is given by: $\mathbf{h}^{k+1}= P_6P_7\mathbf{h}^k,$ with arbitrary $\mathbf{h}^0$.

Just like in the case of \cite{861404}, the Abo-Taleb and Fahmy Algorithm has also been extended by \cite{16592} to design a class of time constrained FIR filters, where certain filter coefficients are set to 0. They denote the set of such indicies by $I_c$. 
The set and its projector are given below:
\begin{equation}
\label{atf3}
C'_{k} \equiv \{\mathbf{a}:\left|D(\omega_k) - \sum_{n=0 , n \not \in I_c}^{(N-1)/2} a(n) \phi_{n}(\omega_k)\right| \leq \lambda(\omega_k)\}
\end{equation}

The projector is given by:
\begin{numcases}{a(n)_{n \not \in I_c}^{i+1}=}
\label{atfp1}
a(n)^i - [\lambda(\omega_k) - |e(\omega_k)|] \text{sign}(e_k) \nonumber\\ .\psi(\omega_k)/ \sum_{m=0, m \not \in I_c}^{(N-1)/2} \psi_m^2 (\omega_k), \text{ if } e(\omega_k) > \lambda(\omega_k)
\nonumber\\
a(n)^i,  \text{ if } e(\omega_k) \leq \lambda(\omega_k)
\end{numcases}

\section{Nyquist filters}
In this section we discuss the Nyquist filter design using POCS. Here we propose an extension of the design methodology used in \cite {861404}. 
A Nyquist filter imposes the following restriction on the impulse response

\begin{numcases}{h(Ln)=}
\label{nfd1}
1/L \text{, for } n=0 \nonumber\\
0 \text{, otherwise }
\end{numcases}

The set formulation for the above constraint is given as $C_{Nyquist}$ and it is easy to check that it is closed and convex.

\begin{eqnarray}
C_{Nyquist} \equiv \{\mathbf{h}:  h \Big(\frac{N-1}{2}\Big) = 1/L \text{ and } h\Big(n- \frac{N-1}{2}\Big)=0  \nonumber\\ \text{ for } \Big(n-\frac{N-1}{2}\Big) \mod  L \equiv 0, \forall n \text{ except } n= (N-1)/2 \}
\end{eqnarray}

\subsection{Projector}
The projection of an arbitrary vector $\mathbf{g} \in \mathbb{R}^M$  on this set is given by

\begin{numcases}{h^*(n)=P_{Nyquist}(g)=}
\label{nfd2}
1/L \text{, if }  n=N-1/2 \nonumber\\
0 \text{, if } \Big(n-\frac{N-1}{2}\Big)\mod  L \equiv 0\nonumber\\ 
g(n) \text{, elsewhere}
\end{numcases}

To prove the above stated result we define
\begin{equation}
I \equiv \Big\{n: \Big(n-\frac{N-1}{2}\Big)\mod  L \equiv 0 \text{ for } n= 0, \ldots N-1 \Big\}
\end{equation}

since our aim is to $\min _{h \in C}\|h-g\|$, it is equivalent to minimizing the following function

\begin{equation}
\label{nfd3}
J=\sum_{n\in I}|h(n)-g(n)|^2 + \sum_{n\in I^c}|h(n)-g(n)|^2
\end{equation}

The definition of the set $C_{Nyquist}$ tells us that we can modify only those $h(n)$ for which $n\in I$, thus its minimum value would be attained if $h(n)=g(n)$ for $n \in I$, which gives us the projector as in (\ref{nfd2}). 
The design algorithm is given by: $\mathbf{h}^{k+1}= P_1P_2P_3P_{Nyquist}\mathbf{h}^k$, with arbitrary $\mathbf{h}^0$.


\section{Simulations}

In this section we show the simulation results for some of the constrained filters described in previous sections. We evaluate the FFT over $M=1024$ discrete frequencies. The iteration stopping criteria is $\|h^{k}-h^{k-1}\|<10^{-6}$. 

\subsection{Example 1}
 As our first example we consider the design of  Linear phase FIR filter with $N=31$, $\alpha=\beta=0.0243$ $ \omega_p=0.4\pi$ and $\omega_s=0.5\pi$. The algorithm converges in 1989 iterations. The POCS design (in blue) is compared with MP design method (dashed line). The results are comparable.

\begin{figure}[!t]
\centering
\includegraphics[width=3.0 in]{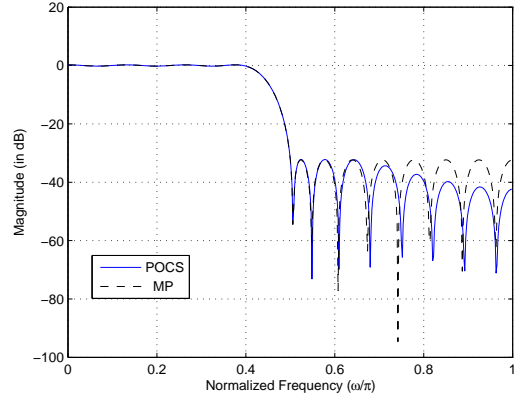}
\caption{Example 1: Linear Phase FIR Low Pass Filter}
\label{fig_sim}
\end{figure}

\subsection{Example 2}
Next the constraint given by (\ref{cons4}) is simulated for the step response of a filter with the following specifications:

\begin{itemize}
\item{$b_1(n)=-0.055$ and $b_2(n)=0.055$ for $n=1, \ldots ,13$ }
\item{$b_1(n)=1-0.055$ and $b_2(n)=1+0.055$ for $n=18, \ldots ,31$ }
\item{$\alpha=\beta=0.13$, the passband and stop-band frequencies are $0.4\pi$ and $0.5\pi$ respectively}
\end{itemize}

\begin{figure*}[!t]
\centering
\subfloat[Step Response]{\includegraphics[width=3.0 in]{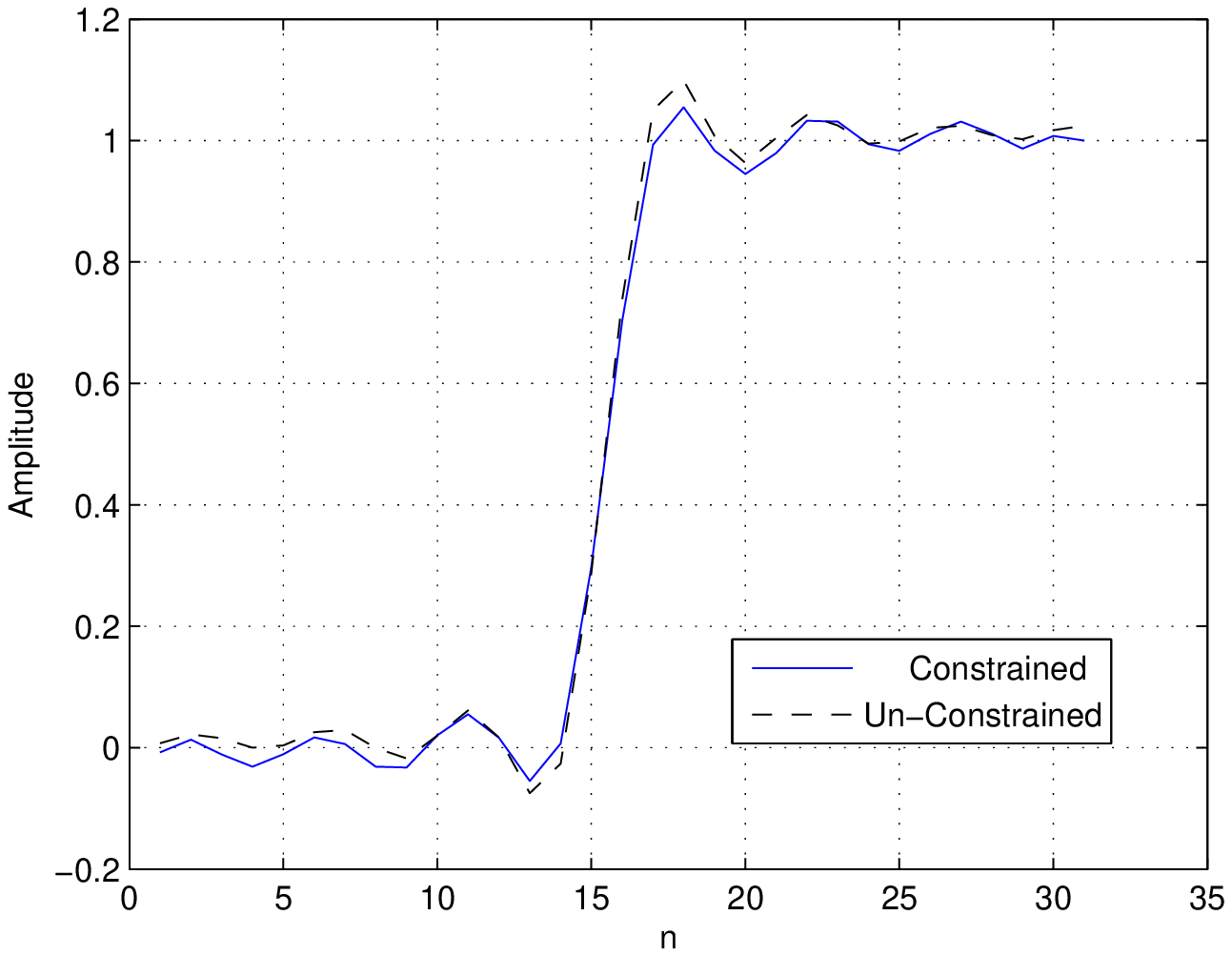}%
\label{conv_1}}
\hfil
\subfloat[Frequency response]{\includegraphics[width=3.0 in]{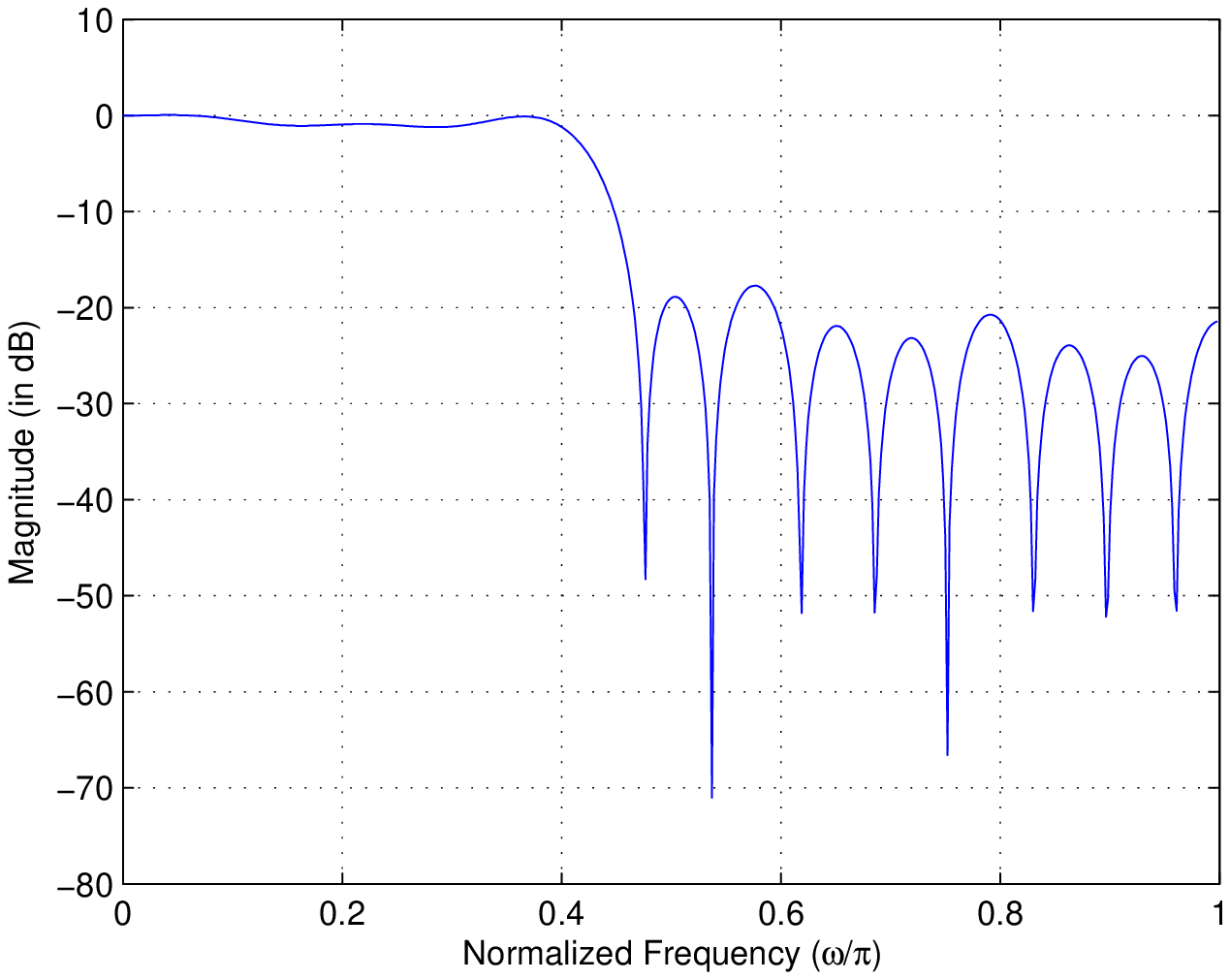}%
\label{conv_2}}
\caption{Example 2: Step Response Constraint}
\label{fig_sim}
\end{figure*}

The difference in the constrained and unconstrained response is visible in  Fig. \ref{conv_1}, however to obtain a constrained step response we must sacrifice the frequency response characteristics as shown in Fig. \ref{conv_2}. We do not alter the response during the amplitude rise.

\subsection{Example 3: Nyquist Filter}
\begin{figure*}[!t]
\centering
\subfloat[Time Domain Response]{\includegraphics[width=3.0 in]{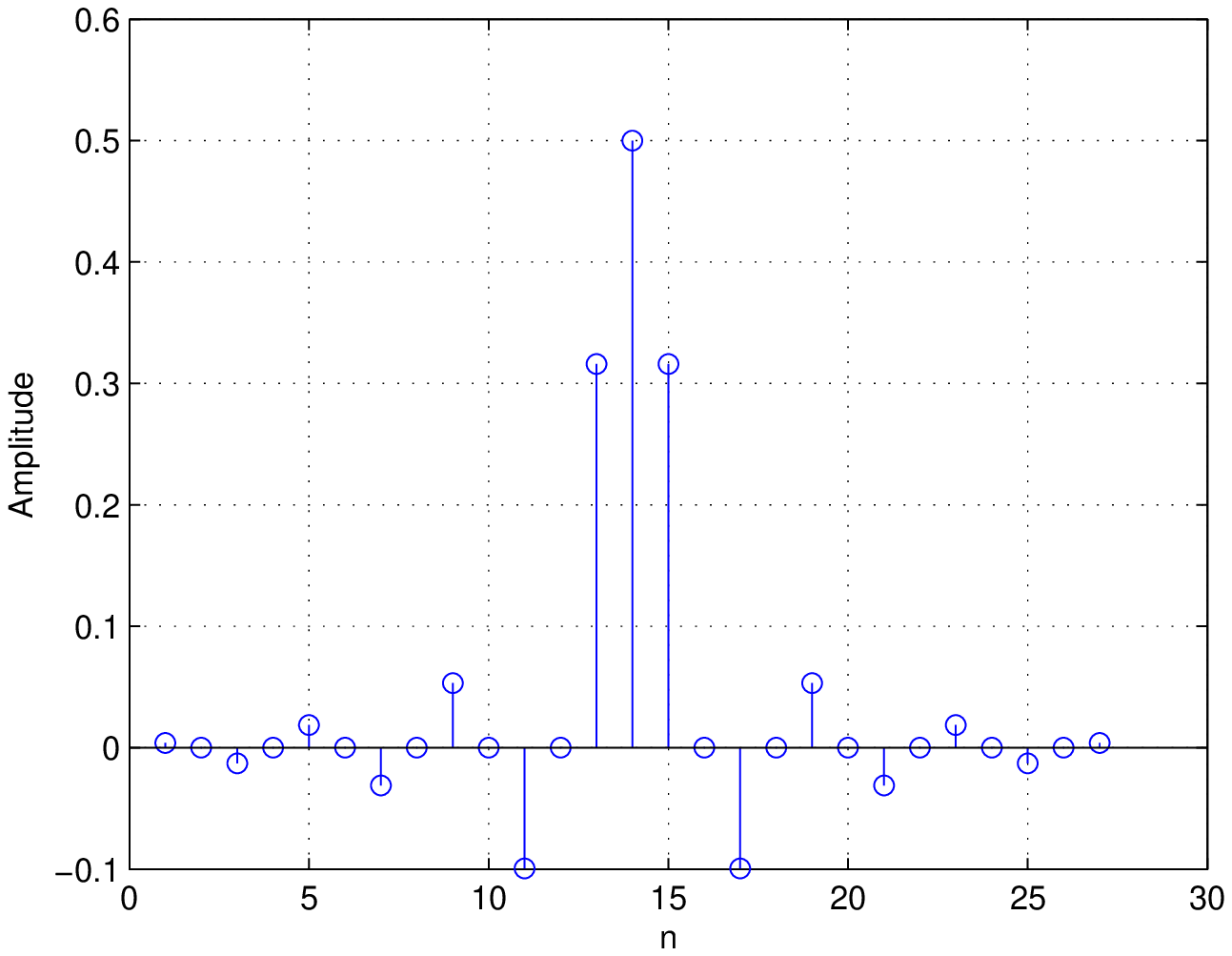}%
\label{hbf_1}}
\hfil
\subfloat[Frequency response]{\includegraphics[width=3.0 in]{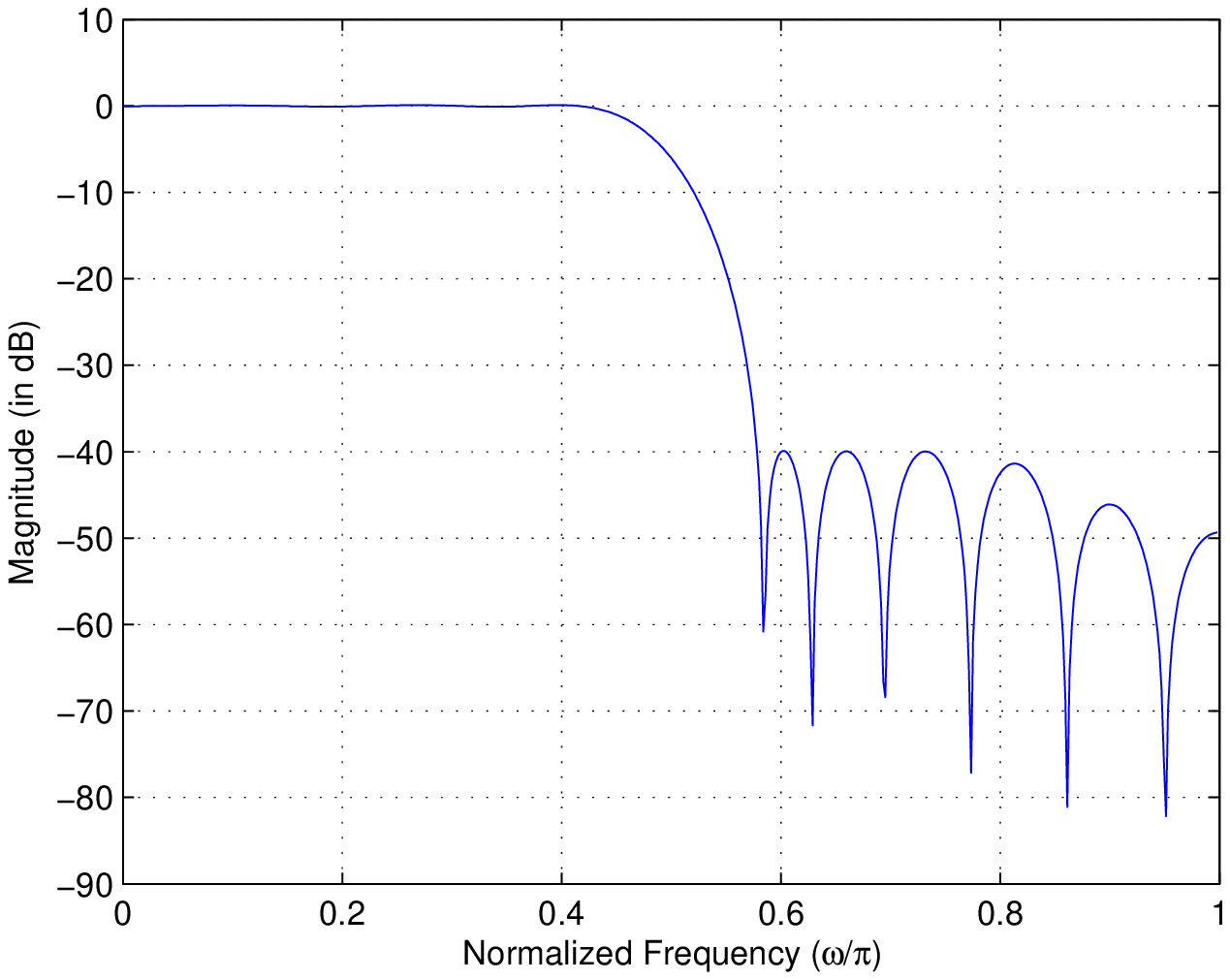}%
\label{hbf_2}}
\caption{Example 3: Half Band Filter}
\label{fig_sim}
\end{figure*}

\begin{figure}[!t]
\centering
\includegraphics[width=3.0 in]{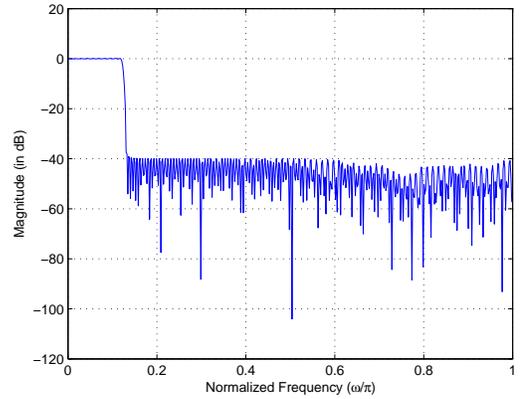}
\caption{Example 3: Nyquist Filter with $N=313 \text{ and }L=8$}
\label{nyq}
\end{figure}

The simulation results for the proposed Nyquist filter design algorithm are presented here. First a Linear phase FIR half band filter design is shown, both in the frequency and the time domain. The design paramters are: $N=27$, $\alpha=\beta=0.01$ $ \omega_p=0.42\pi$ and $\omega_s=0.58\pi$. Evidently, the constraints on the frequency response together with Nyquist filter constraints are satisfied as shown in Fig. \ref{hbf_1} and Fig. \ref{hbf_2}. Next we designed a filter of order, $N= 313$ and $L=8$ for the following design parameters: $\alpha=\beta=0.01$ $ \omega_p=0.3731$ and $\omega_s=0.4123$ Fig. \ref{nyq}.

\subsection{Example 4: 2-D Filters}
The last example we consider is that of simulating 2-D filters, which serves to illustrate the fact that POCS methods are easily extended to $m$ dimensional case. We design a linear phase two fold symmetric circular FIR filter wih the following parameters: $M=128, N=17, \alpha=0.05, \beta=0.05, \omega_p=0.43\pi, \text{ and }\omega_s=0.63\pi$.

\begin{figure}[!t]
\centering
\includegraphics[width=3.0 in]{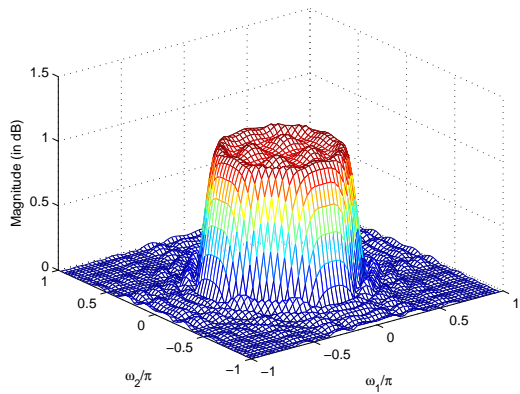}
\caption{Example 4: 2-D Filter}
\label{md}
\end{figure}

\section{Conclusion}
In this paper, we reviewed convex set formulations of the constraints for  linear phase FIR filter design problem, we then extended the design by incorporating additional constraints for Nyquist filters. Throughout the paper,  we have maintained an intutive approach to the POCS methods for easy and clear understanding. In particular we have proposed a simple design methodology for Nyquist filter. Though the algorithm does not guarantee optimal solution, the resulting design satisfies all of the required constraints.

\ifCLASSOPTIONcaptionsoff
  \newpage
\fi




\bibliographystyle{IEEEtran}
%



%





\end{document}